# Gates controlled parallel-coupled double quantum dot on both single layer and bilayer graphene

*Lin-Jun Wang $^§$, Guo-Ping Guo $^{§*}$, Da Wei $^§$, Gang Cao $^§$, Tao Tu $^{§*}$, Ming Xiao $^§$, Guang-Can Guo $^§$*

§ Key Laboratory of Quantum Information, University of Science and Technology of China, Chinese Academy of Sciences, Hefei 230026, People's Republic of China

*A. M. Chang$^†$*

† Department of Physics, Duke University, Durham, NC 27708, USA.

\* Corresponding authors. Email: gpguo@ustc.edu.cn, tutao@ustc.edu.cn

**ABSTRACT:** Here we report the fabrication and quantum transport measurements of gates controlled parallel-coupled double quantum dot on both bilayer and single layer graphene. It is shown that the interdot coupling strength of the parallel double dots can be effectively tuned from weak to strong regime by both the in-plane plunger gates and back gate. All the relevant energy scales and parameters of the graphene parallel-coupled double dot can be extracted from the honeycomb charge stability diagrams revealed through the transport measurements.

Gate controlled double quantum dot (DQD) system has been considered as a promising candidate of spin-based solid state qubits for the quantum computation processing.[1,2] Many efforts and progresses have been made in the double dot devices research of various materials, including GaAs two-dimensional electron gas,[3,4] semiconductor nanowires,[5,6] and carbon nanotube.[7–9] The nature two-dimensional material, graphene, has attracted extensive interest due to its distinguishing electronic quality and flexibility in device designs.[10–17] In addition, it is theoretically predicted that spin decoherence time in graphene can be much long due to its weak spin-orbit coupling and largely absent hyperfine interaction, which has significant meaning for spin-based quantum processors. Although various nano-devices including double quantum dot in series have been developed in graphene, there is no report about parallel-coupled graphene double quantum dot to our knowledge. Compared to DQD in series, the parallel DQD is an ideal artificial system for investigating the interaction and interference. Rich physical phenomena, such as Aharonov- Bohm (AB) effect, Kondo regimes and Fano effect, can be observed in parallel DQD.[18–23] As a example of applications of parallel DQD, probing and controlling the bonds of an artificial molecule has been demonstrated in parallel DQD in GaAs as the parallel access to the two dots enables correlated tunneling of two valence electrons simultaneously.[24]

Here we report a design and fabrication of double dot structure coupled in parallel on both bilayer and single layer graphene flake, which may open a door for us to study the rich parallel DQD physical phenomena in the peculiar new material, graphene. It is shown that the parallel graphene structure can be tuned from a strong-coupling resulted artificial molecule state to weakly coupled two dots by adjusting of three in-plane plunger gates and a global back gate. All the relevant energy scales and parameters are extracted from the quantum transport charge stability diagrams of this parallel coupled bilayer graphene double dot system. As shown in Figure 1(a), the central islands (dots), nano-constrictions (forming tunnel barriers) and in-plane plunger gates are all carved from the entire two-dimensional (2D) sheet. The bridge plunger gate is separated from the drain part of graphene by a layer of over exposed PMMA. This design and fabrication method , benefitting from the two dimensional structure and high crystal and electronic quality of graphene[11-13], is different from the way on the GaAs semiconductor and nanowire devices, [3-9] and give us a new perspective in the future integrated nano-devices.

Graphene flakes are produced by mechanical cleaving of bulk graphite crystallites by Scotch tape.[25] By using highly doped Si substrate with 100nm thick $SiO_2$ on top, we can identify the monolayer, bilayer and few layer graphene through optical microscope. The single layer graphene and bilayer graphene were further check by Raman spectrum. The results presented here are based on the measurement of one of our devices made on a bilayer grpahene flake. We first make an array of gold markers on the substrate before the graphene flakes are transferred to, these markers will help us locate graphene pieces in the following nanofabrication steps. After the graphene flake was transferred, a 50nm thick layer of polymethyl methacrylate (PMMA) is spun for the electron beam lithography (EBL) to form a designed pattern, followed by reactive ion etching processing in an $O_2$/Ar (50:50) plasma, during which unprotected graphene part will be removed; the pattern of PMMA are then transferred to the graphene flake. We again spin a layer of PMMA which will be over exposed[23, 26] to form a bridge for the supporting of plunger gate GM that needs to be separated from the drain part of graphene---please see Figure 1(b). The final step is to make the metal contacts, which are defined by standard EBL, followed by the E-beam evaporation of Ti/Au (2nm/50nm).

Figure 1(a) shows a scanning electron microscope (SEM) image of one sample with the same structure as the device we measured. Two central islands with diameter of 100 nm connect through 30nm wide narrow constrictions to the source and drain regions. Another narrow constriction (35nm in both width and length) connects the two central islands. Seven in-plane plunger gates GL, GR, GM, PSL, PDL, PSR, PDR are integrated in close proximity to the dots. GL, GR, GM are respectively designed to adjust the energy level of left dot, right dot, inter-dot coupling strength. And PSL, PDL ( PSR, PDR) are for the tuning of the coupling of the left (right) dot to source and drain. The n-type heavily doped silicon substrate is used as a global back gate. Figure 1(b) shows a schematic picture of the device. All the devices were primarily tested to check the functionality of all the gates in a liquid helium storage dewar at 4.2K. Then the samples were mounted on a dilution refrigerator equipped with filtering wirings and low-noise electronics at the base temperature of 10mK. To maintain consistency, we will use the data from one sample only in the following.

Figure 2(a) shows the conductance through the double dot as a function of back gate using a small AC source-drain excitation voltage of 20μV rms. From the overall shape of the conductance we clearly see that the conductance is strongly suppressed from the back gate interval -2V to 8V. It is estimated that the neutral point (NP) would be a positive value in this region, showing the p-doing characteristic of the graphene flake. However the exact position cannot be determined. This shift of NP position from zero is attributed to an unintentional doping of the films by absorbed water.[24] The upper inset in figure 2(a) zooms into the interval from-2V~8V. We can see several non-periodic coulomb blockade peaks due to the non-single dot behavior. Figure 2(b) shows the coulomb diamonds as a function of voltage applied on the left plunger gate $V_{GL}$ and source-drain DC bias $V_{SD}$. Other gates are set to $V_{BG}$=3.5V, $V_{GR}$=0V, $V_{GM}$=0V. The electron temperature is estimated to be around 120mk from the temperature dependence of the CB peaks width.

Figure 3(a) shows color scale plot of the measured conductance of the double dot as a function of $V_{GL}$ and $V_{GR}$ detected in standard ac lock-in technique with an excitation ac voltage 20μV at frequency 11.3Hz. A dc bias of 0.3mv is applied, the back gate voltage $V_{BG}$ is fixed at 5V and the middle plunger gate $V_{GM}$ is -0.45V. The hexagon pattern characteristic for double dot coupled in parallel is clearly visible. From the model of purely capacitive coupled dots as illuminated by Figure 3(d), the energy scales of the system can be extracted.[3,12,13,15] The capacitance of the dot to the side gate can be determined from measuring the size of the honeycomb as shown in Fig. 3(a) (b), $\Delta V_{GL}$=0.087V, $\Delta V_{GR}$=0.053V, $\Delta V^m_{GL}$=0.0261V, $\Delta V^m_{GR}$=0.0133V, therefore $C_{GL} = e/\Delta V_{GL} = 1.84aF$, $C_{GR} = e/\Delta V_{GR} = 3.0aF$. With a large DC bias of 0.3mV, we can get $\delta V_{GL}$=0.013V, $\delta V_{GR}$=0.01V as shown in figure 3(c). The lever arm between the left (right) gate $V_{GL}$ and the left (right) dot can be calculated as $\alpha_{GL}=V_{bias}/\delta V_{GL}$=0.023 ($\alpha_{GR}=V_{bias}/\delta V_{GR}$=0.03). The total capacitances of the dots can then be calcualted as $C_L=C_{GL}/\alpha_L$=79.8aF, $C_R=C_{GR}/\alpha_R$=100.4aF, the corresponding charging energy $E_{CL}=\alpha_{GL}\cdot\Delta V_{GL}$=2.0mev, $E_{CR}=\alpha_{GR}\cdot\Delta V_{GR}$=1.6mev, the coupling energy between the two dots $E_{CM}=\alpha_{GL}\cdot\Delta V^m_{GL}$=0.3meV. It is also noted that the lever arms between the left gate and the right dot and vice versa can be determined from the slope of the cotunneling lines delimiting the hexagons. These crossing couplings only modify the results very slightly and are neglected usually.[12,13,15]

Back gates are very common utilized in graphene and nanowire based nanodevices. In graphene, gate gate can effectively tune the Fermi level of the material to change the carriers between holes and electrons. Here we demonstrate an evolution of conductance patterns, which indicates the stability diagram changes from weak to strong tunneling regimes in different back gate regimes. Figure 4(a)-(c) successively represent the weak, medium and strong coupling of the two dots. (a) $V_{BG}$= -4V, $V_{GM}$= 0V, Bias= -0.6mV, (b) $V_{BG}$= 3V, $V_{GM}$=0.6V, Bias=-1mV, (c) $V_{BG}$= -3V, $V_{GM}$= 0V, Bias= -1mV. In the case of weak coupling regime as shown in figure 4(a), the two dots can be treated independently except for coupling by capacitance. While in strong coupling regime, figure 4(c), the two dots are interacting with each other through the large quantum mechanical tunnel coupling, analogous to a two-atom molecule with covalent bonding, behave like a single dot. In the middle coupling region as shown in figure 4(b), the two dots have both capacitance and tunnel couplings. Figure 4(a'), (b'), (c') show the sketches of the characteristic of two dots' bonding states.

We also study how the middle plunger gate GM affects the interdot coupling. Figures 5(a) (b) and (c) show the charge stability diagram for three different regions of coupling strength. [(a) weak, (b) medium, (c) strong]. In these measurements, back gate voltage $V_{BG}$=3V, Source-Drain DC bias is set to -1mv, the scan region of GL and GL are the same. Only the voltage applied to the gate GM is adjusted as (a) $V_{GM}$=-0.15V, (b) $V_{GM}$=-0.2V and (c) $V_{GM}$=0.45 V. The corresponding coupling energy between the dots are (a): $E_{CM}$=0.58meV, (b): $E_{CM}$ =1.34 meV and (c): $E_{CM}$ =4.07 meV. Figure 5(d) indicates the coupling energy changes with the gate voltage $V_{GM}$. As in the previous reports of graphene DQD in series,[12, 15] the interdot coupling is nonmonotonicly depended on the applied gate voltage. Although the detailed reasons for this non-monotony are undetermined, we assumed that one key factor will be the disorders in graphene introduced by either fabrication steps or substrate.[27] Many more efforts are still needed to address this issue for the realization of practical graphene based nanodevices.

Our device is very stable temporally. The data can readily be repeated after several days in the dilution refrigerator. In addition, we have designed and fabricated another new structure of a parallel-coupled DQD integrated with two quantum point contact sensors (QPC) in single layer graphene, as shown in figure 6(b) (c). For this new structure, we can get similar charge stability diagram of the parallel DQD as figure 6(a), by the direct quantum transport tests even at 4.2K. From these primary tests, no remarkable difference is founded between PDQD in bilayer and single layer graphene. The expected energy gap in bilayer grapehen may have been enshrouded in size effects. These results show that the present method of designing and fabricating graphene parallel DQD is general and reliable. Both bilayer and single layer graphene can be exploited in this application. The further detailed experiments include exploiting QPC as charge sensor for graphene parallel DQD are still in processing.

In conclusion, we have reported a design and fabrication of gate-controlled parallel-coupled double-dot both bilayer graphene. It is also shown that the interdot coupling, as well as the transmission of the constrictions connecting the dots to the leads, can be largely tuned by graphene in-plane gates. With the quantum transport honeycomb charge stability diagrams, a common model of purely capacitively coupled double dot is used to extract all the relevant energy scales and parameters of bilayer graphene parallel-coupled double-dot. Although many more effects are still needed to further upgrade and exploit this new designed graphene structure, the present results have intensively demonstrated the promise of the realization of graphene nanodevice and desirable study of rich parallel DQD physical phenomena in graphene.

**Acknowledgements:** This work was supported by the National Basic Research Program of China (Grants No. 2011CBA00200) and the National Natural Science Foundation of China (Grants No. 10804104, No. 10874163, No. 10934006, No. 11074243). Work at Duke was supported in part by NSF-DMR 0701948.

**FIGURE CAPTIONS**

**FIGURE 1.** Parallel double quantum dot (DQD) device configuration. (a) Scanning electron microscope image of the designed parallel-coupled graphene double dot structure in this work. The diameters of the two dots are both 100nm, the constriction between the two dots is 35nm in width and length. The four narrow parts connecting the dot to source and drain parts have a width of 30nm. Seven in-plane plunger gates GL, GR, GM, PSL, PDL, PSR and PDR are respectively designed to adjust the energy level of left dot, right dot, and inter-dot coupling strength, the tuning of the coupling of the left dot to source and drain, the tuning of the coupling of the right dot to source and drain. The n-type heavily doped silicon substrate is used as a global back gate. (b) Schematic picture of the device. We use a layer of overexposed PMMA to create a bridge for the gate GM to get separated from the drain part of graphene.

**FIGURE 2.** Transport measurements of parallel DQD. (a) Conductance through the double dot as a function of back gate voltage $V_{BG}$. The upper inset zooms in the low-G region -2~8V, revealing several non-periodic coulomb peaks. (b) Color plot of coulomb diamonds: differential conductance $G_{diff}=dI/dV$ as a function of voltage applied on the left plunger gate $V_{GL}$ and source-drain DC bias $V_{SD}$. Other gates voltage: $V_{BG}=3.5V$, $V_{GR}=0V$, $V_{GM}=0V$. A 20μV of ac excitation voltage is applied between the source and drain electrodes for standard lock-in measurement.

**FIGURE 3.** Charge stability diagrams of the parallel-coupled DQD. (a) Parallel DQD conductance as a function of plunger gate voltage $V_{GL}$ and $V_{GR}$. Back-gate voltage is fixed at $V_{BG}=5V$, middle plunger gate $V_{GM}$ is -0.45V, and a 20μV of ac excitation and a 0.3mV of dc bias are applied to the source-drain electrodes. The red dash lines are guides to the eyes showing the honeycomb pattern. (N, M) represents the carriers in the left and right dot respectively. (b) Zoom-in of the area (N, M) of the honeycomb pattern. The relevant parameters can be extracted as $\Delta V_{GL}=0.087V$, $\Delta V_{GR}=0.053V$, $\Delta V^m_{GL}=0.0261V$, $\Delta V^m_{GR}=0.0133V$. (c) Zoom-in of a vertex pair with white dash lines. We get $\delta V_{GL}=0.013V$, $\delta V_{GR}=0.01V$ from the image. (d) Capacitance model for the analysis of the double dot system.

**FIGURE 4.** Interdot coupling changes with the middle gate voltage $V_{BG}$. (a) weak coupling regime, $V_{BG}$= -4V, $V_{GM}$ = 0V, Bias= -0.6mV, (b) medium coupling regime, $V_{BG}$= 3V, $V_{GM}$=0.6V, Bias=-1mV, (c) strong coupling regime, $V_{BG}$= -3V, $V_{GM}$= 0V, Bias= -1mV. (a$^{'}$), (b$^{'}$), (c$^{'}$) show the sketches of the characteristic of two dots' bonding states. In (a$^{'}$) two dots are more like separated ones besides

capacitance coupling. In (b') the two dots have both capacitance and tunnel coupling. In (c') strong tunneling coupling make the two dots merged into one dot.

**FIGURE 5.** Stability diagrams showing the influence of middle gate voltage $V_{GM}$ on the dots coupling. All parameters are kept the same: back gate voltage $V_{BG}$=3V, Source-Drain DC bias is set to -1mv, the scan region of GL and GL are the same. (a) weak coupling regime, $V_{GM}$=-0.15V, (b) medium coupling regime, $V_{GM}$=-0.2V (c) strong coupling regime, $V_{GM}$=0.45 V. (d) coupling energy as a function of the middle gate voltage $V_{GM}$. A, B, C point here represent the corresponding coupling energy in (a), (b), (c).

**FIGURE 6.** Device structure and characterization in the CB regime of another parallel DQD with QPCs based on single layer graphene (a) Characteristic honeycomb structure of the conductance through the parallel DQD as a function of two in-plane plunger gates $V_{GL}$ and $V_{GR}$, which is revealed by DQD direct transport measurement at 4.2K (b) Scanning electron microscope image of the designed parallel graphene DQD integrated with two QPCs device. (c) Schematic picture of the device.

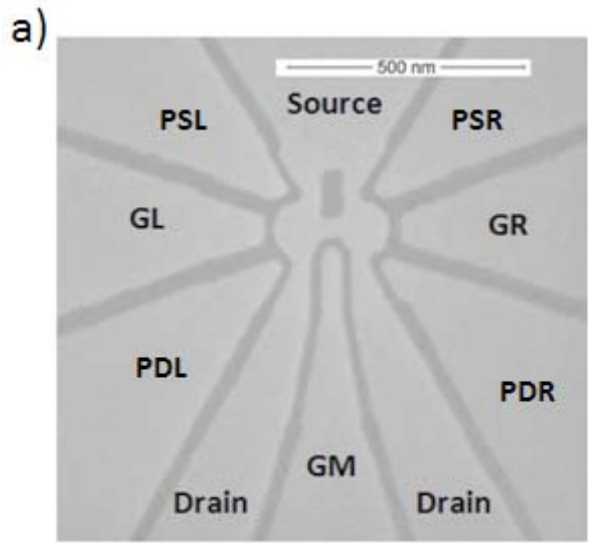 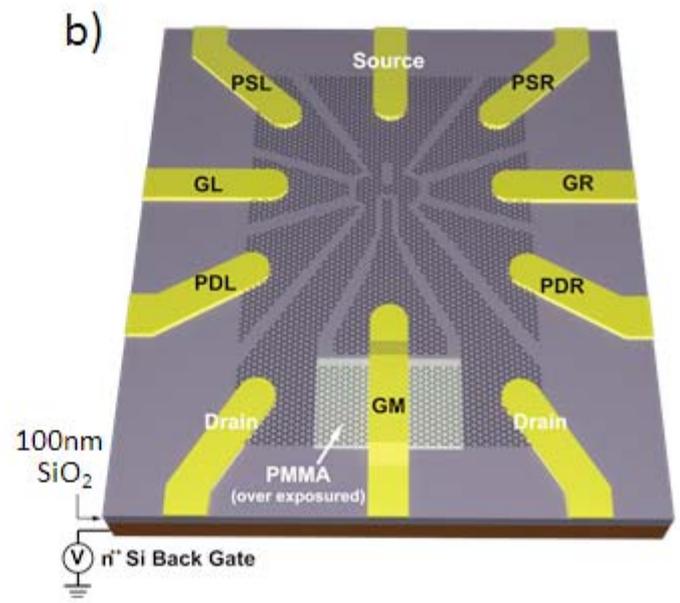

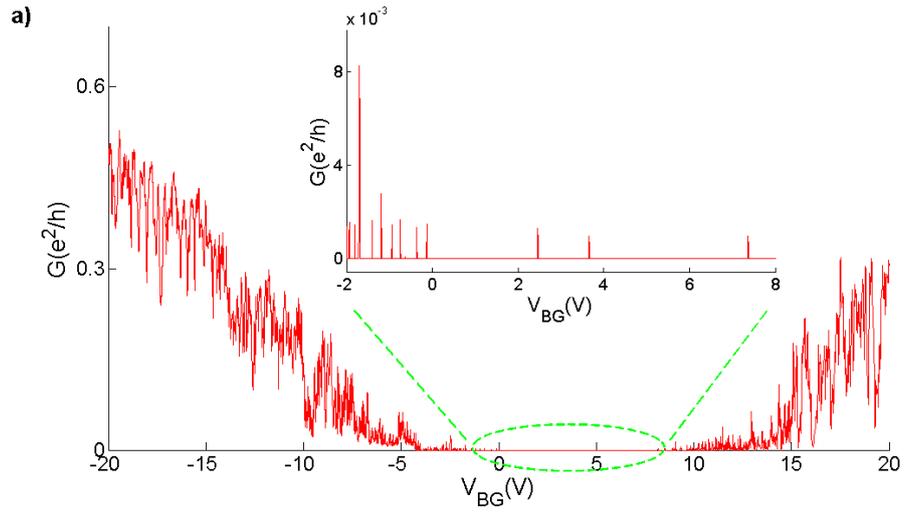

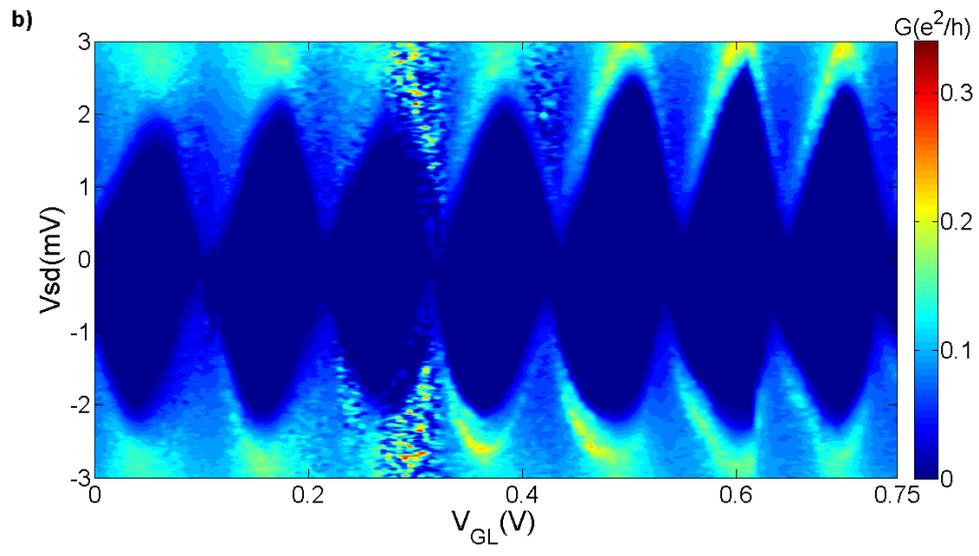

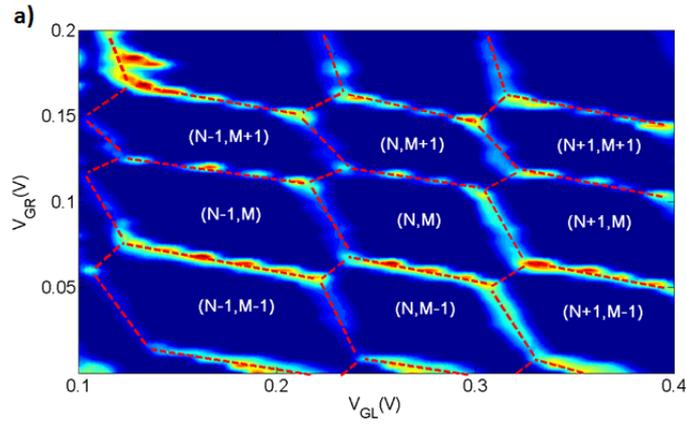
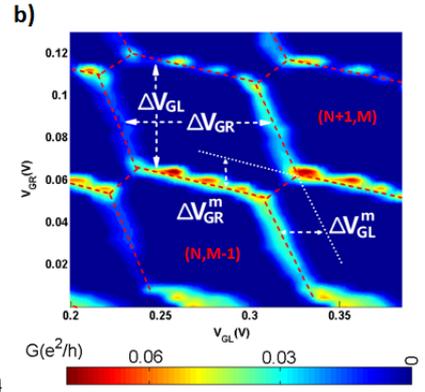
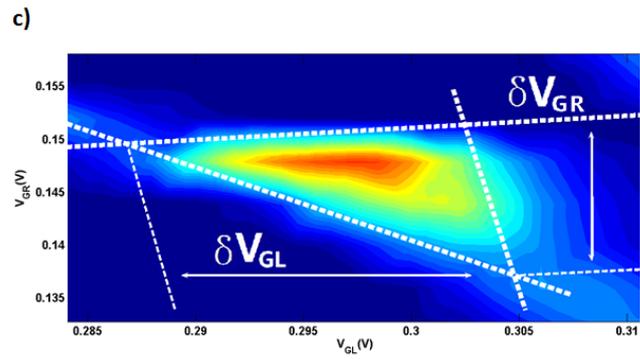
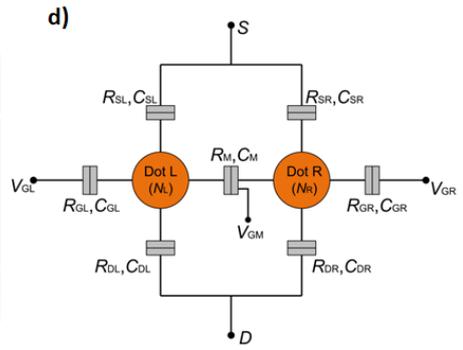

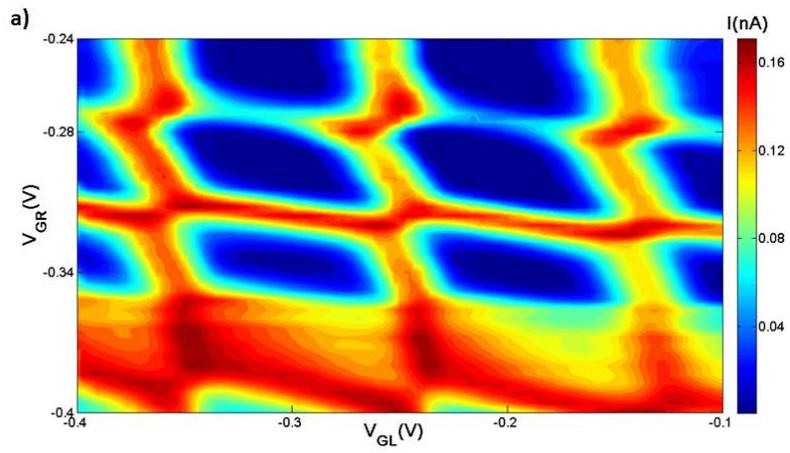
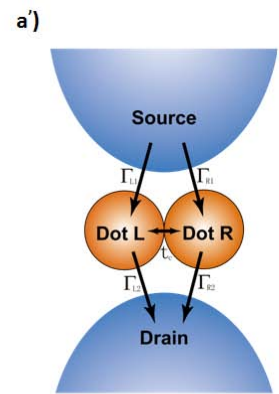

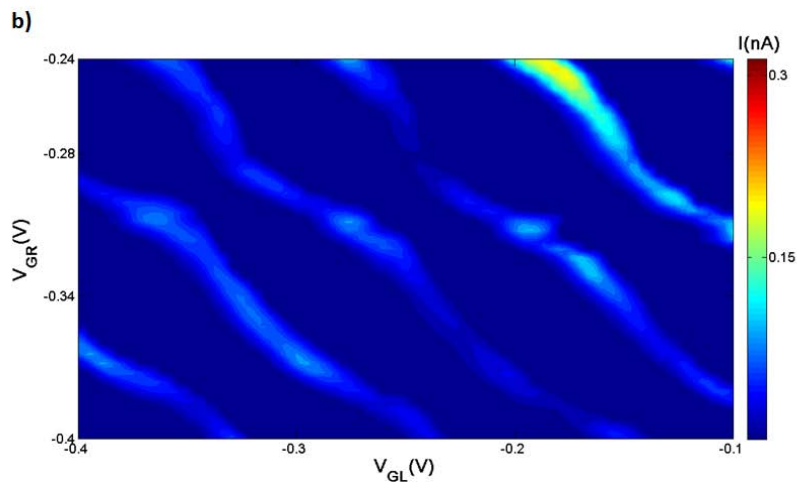
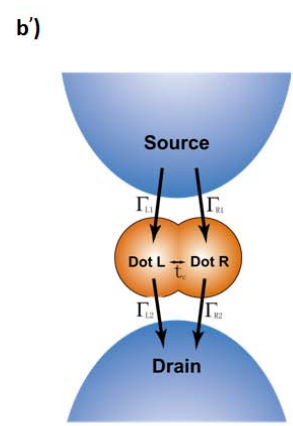

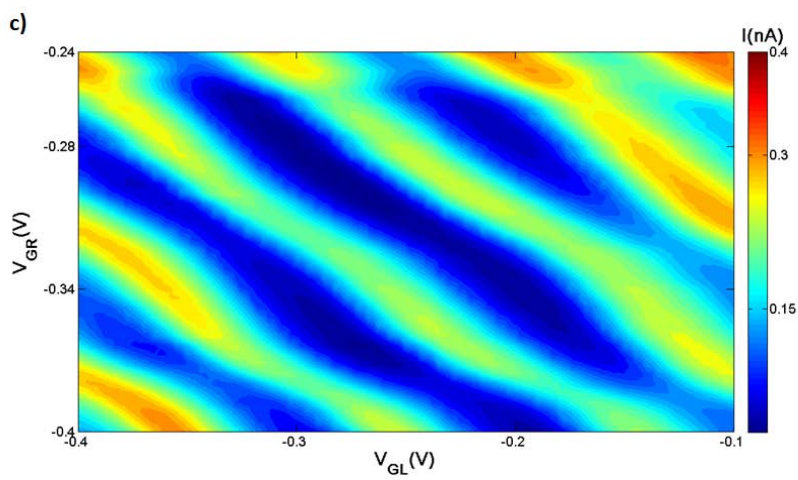
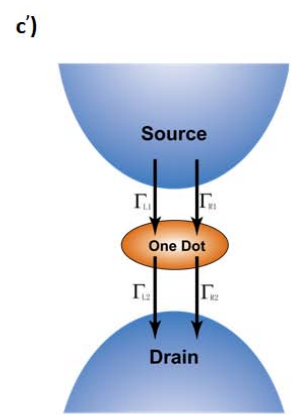

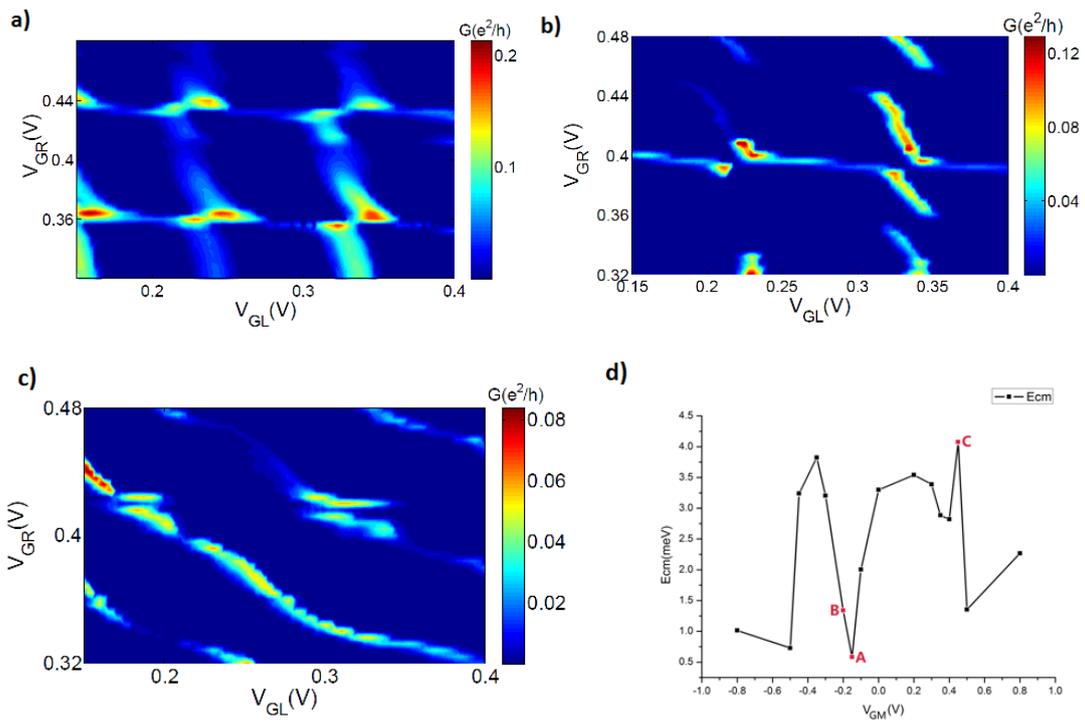

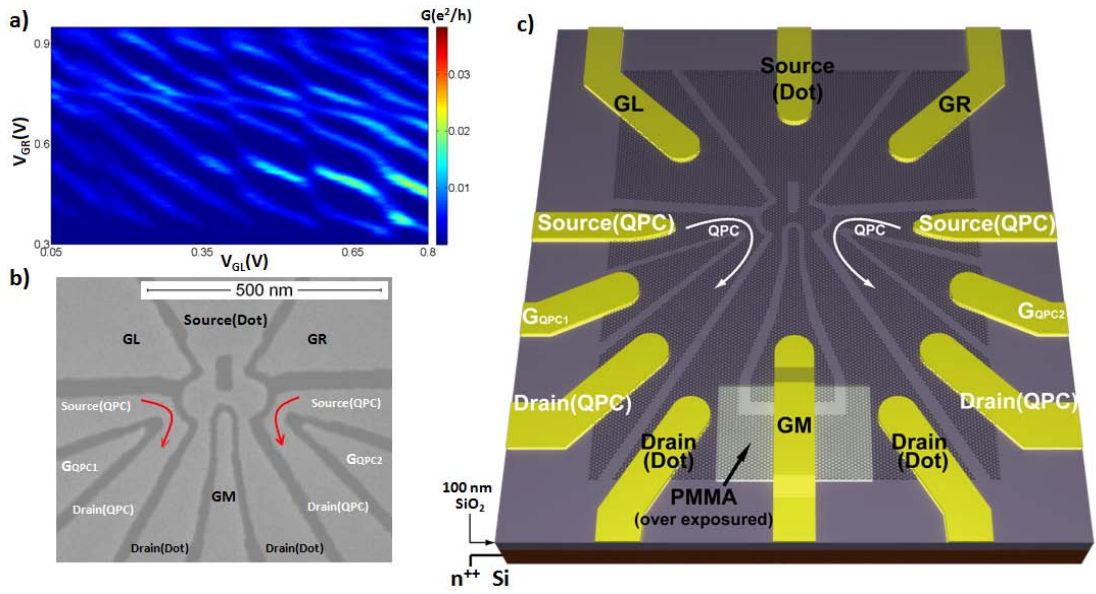